\documentclass{article}

\usepackage[english]{babel}
\usepackage[T1]{fontenc}
\usepackage[dvips]{epsfig} 
\usepackage[cmex10]{amsmath} 
\usepackage{amssymb}
\usepackage[tight, footnotesize]{subfigure}
\usepackage{url}
\usepackage{bm}
\usepackage{subfigure}
\usepackage{theorem}
\usepackage{color}
\usepackage{fancybox}

\usepackage{shadow}
\usepackage{float}
\usepackage{comment}
\usepackage{graphicx}
\usepackage{psfrag}
\usepackage{IEEEtrantools}

\newtheorem{hypothesis}{Assumption}

\newtheorem{proposition}{Proposition}

\newenvironment{system}[1][rCL]
{\left\lbrace\begin{IEEEeqnarraybox}[][c]{#1}}
{\end{IEEEeqnarraybox}\right.}

\newcommand{\rot}{\mathrm{rot}}

\newcommand{\RR}{\mathbb{R}}

\graphicspath{figures/}

\title{Nonlinear Feedback Control of Axisymmetric Aerial Vehicles}

\author{Daniele Pucci$^{1}$, Tarek Hamel$^2$, Pascal Morin$^3$\thanks{Corresponding author}, Claude Samson$^{4}$ \\~\\
$1$ Italian Institute of Technology, Genova, Italy, daniele.pucci@iit.it\\
$2$ I3S/UNSA, Sophia-Antipolis, France, thamel@i3s.unice.fr \\
$3$ ISIR-UPMC, Paris, France, morin@isir.upmc.fr \\
$4$ INRIA, I3S/UNSA, Sophia Antipolis, France, claude.samson@inria.fr, csamson@i3s.unice.fr}

\begin{document}


\maketitle


\begin{abstract}
We investigate the use of simple aerodynamic models for the feedback control of aerial vehicles with 
large flight envelopes. Thrust-propelled vehicles with a body shape symmetric with respect to the thrust axis are considered.
Upon a condition on the aerodynamic characteristics of the vehicle, we show that the equilibrium orientation
can be explicitly determined as a function of the desired flight velocity. This allows for the adaptation of previously
proposed control design
approaches based on the thrust direction control paradigm. Simulation results conducted by using measured aerodynamic characteristics of quasi-axisymmetric bodies illustrate the soundness of the proposed approach.
\end{abstract}


\section{Introduction}
Alike other engineering fields, flight control makes extensive use of linear control techniques \cite{2003_STEVENS}. One reason is the 
existence of numerous tools to assess the robustness properties of a linear feedback controller \cite{rdb12} (gain margin, phase margin, 
$H_2$, $H_\infty$, or LMI techniques, etc.). Another reason is that flight control techniques have been developed primarily for full-size 
commercial airplanes that are designed and optimized to fly along very specific trajectories (trim trajectories with a very narrow 
range of angles  of attack). Control design is then typically achieved from the linearized equations of motion along desired trajectories. 
However, some aerial vehicles are required to fly in very diverse conditions that involve
large and rapid variations of the angle of attack. Examples are given by fighter aircraft, convertible aircraft, or small 
Unmanned Aerial vehicles (UAVs) operating in windy environments. As a matter of fact, some Vertical Take-Off and Landing (VTOL) vehicles, like e.g.
ducted fans, are often subjected to large variations of the angle of attack when transitioning from hover to horizontal cruising flight.
It then matters to ensure large stability domains that are achievable via the use of nonlinear feedback designs.

Nonlinear feedback control of aircraft can be traced back to the early eighties. Following  \cite{ss80}, control laws based on the 
dynamic inversion technique have been proposed to extend the flight envelope of military aircraft (see, e.g., \cite{ws05} and the 
references therein). The control design strongly relies on tabulated models of aerodynamic forces and moments, like the 
High-Incidence Research Model (HIRM) of the Group for Aeronautical Research and Technology in Europe (GARTEUR) \cite{mu97}. 
Compared to linear techniques, this type of approach allows one
to extend the flight domain without involving gain scheduling strategies. The angle of attack is assumed to remain away from the
stall zone. However, should this assumption be violated the system's behavior is unpredictable. Comparatively, nonlinear feedback 
control of VTOL vehicles is more recent, but it has been addressed with a larger variety of techniques. 
Besides dynamic inversion \cite{hsm92}, other techniques include Lyapunov-based design \cite{2002_MARCONI,2003_ISIDORI}, 
Backstepping \cite{Bouabdalla05}, Sliding modes \cite{Bouabdalla05,Xu08}, and Predictive control \cite{kim02,bph07-ifac}. 
A more complete bibliography on this topic can be found in \cite{hhms13}. Since most of these studies address the stabilization 
of hover flight or low-velocity trajectories, little attention has been paid
to aerodynamic effects. These  are typically either ignored or modeled as a simple additive perturbation, the effect of which has to be 
compensated for by the feedback action. In highly dynamic flight or harsh wind conditions, however, aerodynamic effects become important.
This raises several questions, seldom addressed so far by the control and robotics communities, such as, e.g., 
{\em which models of aerodynamic effects should be considered for the control design?} or {\em which feedback control solutions
can be inferred from these models so as to ensure large stability domains and robustness?} 

Classical methods used in aerodynamic modelling to precisely describe aerodynamic forces, e.g. computational fluid dynamics 
(CFD) or wind tunnel measurements, do not provide analytical
expressions of aerodynamic characteristics.  From a control design perspective they are useful to finely tune a controller around a 
given flight velocity, but exploiting them in the case of large flight envelopes (i.e., that involve strong variations of
either the flight velocity or the angle of attack) is difficult. In this paper we advocate the use of simple analytical models of aerodynamic
characteristics. Although relatively imprecise, these models may account for important structural properties of the system in a large
flight envelope. The main idea is to exploit these properties at the control design level and rely on the robustness 
of feedback controllers to cope with discrepancies between the model and the true aerodynamic characteristics. More precisely, for the class
of vehicles with a body-shape symmetric w.r.t. the thrust axis, we provide conditions on the aerodynamic coefficients under which the
vehicle's equilibrium orientation  associated with a desired flight velocity is explicitly (and uniquely) defined. We also
show that such conditions are satisfied by simple models that approximate at the first order aerodynamic characteristics of real systems
reported in the literature. The control design then essentially consists in aligning the thrust direction with the desired 
equilibrium orientation and monitoring the thrust intensity to compensate for the intensity of external forces. 
This corresponds to the thrust direction control paradigm, which has been exploited for VTOL vehicles either by neglecting aerodynamic 
effects \cite{ghm05}, or by considering systems submitted to drag forces only  \cite{hhms09}. Although the determination
of the vehicle's equilibrium orientation is straightforward in these cases, this is a major issue for more general vehicles (see \cite{pucciPhd} for
more details). By showing that the thrust direction control paradigm 
can be extended to aerial vehicles submitted to significant lift forces, this paper makes a step towards a unified control 
approach for both VTOL vehicles and airplanes.

The paper is organized as follows. Section \ref{sec-background} provides the notation and background.
In Section \ref{sec:spheEquiv}, we show that for a class of symmetric bodies the dynamical equations of motion can be transformed into 
a simpler form that allows one to explicitly determine the equilibrium orientation associated with a desired flight velocity. 
This transformation is then used in Section \ref{sec-control} to propose a feedback control design method applicable to several vehicles 
of interest.

\section{Notation and background}
\label{sec-background}
Throughout the paper, ${\bm E}^3$ denotes the $3D$ Euclidean vector space and vectors in ${\bm E}^3$ are denoted with bold letters.
Inner and cross products in ${\bm E}^3$ are denoted by the symbols $\cdot$ and $\times$ respectively.
 
\begin{figure}[htbp]
\centering
\def\svgwidth{.4\linewidth}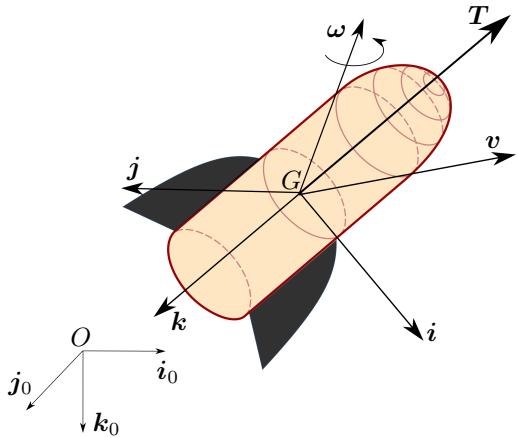
\caption{Notation.}\label{figNotation}
\end{figure}

Let $\mathcal{I} = \{O;\bm{i}_0,\bm{j}_0,\bm{k}_0\}$ denote a 
fixed inertial frame with respect to (w.r.t.) which the vehicle's absolute pose is measured (see Figure~\ref{figNotation}). 
This frame is chosen as the NED frame (North-East-Down) with $\bm{i}_0$ pointing to the North, $\bm{j}_0$ pointing to the East, 
and $\bm{k}_0$ pointing to the center of the Earth. Let $\mathcal{B} = \{G;\bm{i},\bm{j},\bm{k}\}$ denote a frame attached to the 
body, with $G$ the body's center of mass. The linear and angular velocities $\bm{v}$ and $\bm{\omega}$ of the body frame $\mathcal{B}$ are 
then defined by
\begin{equation}
  \label{eq:kinematics}
  \bm{v} :=\frac{d}{dt}\bm{OG}\ , 
  \quad \quad
  \frac{d}{dt} (\bm{i},\bm{j},\bm{k}) := \bm{\omega} \times (\bm{i},\bm{j},\bm{k}) \ ,
\end{equation}
where, here and throughout the paper, the time-derivative is taken w.r.t. the inertial frame $\mathcal{I}$.

\subsection{Equations of motion}
\label{equationsMotion}
Let $\bm{F}$ and $\bm{M}$ denote respectively the resultant of control and external forces acting on a rigid body of mass $m$ and the
moment of these forces about the body's center of mass $G$. 
Newton's and Euler's theorems of Mechanics state that
\begin{equation}
  \label{eq:newtonEuler}
  \dot{\bm{q}} = \bm{F}\ , 
  \quad \quad
  \dot{\bm{h}} = \bm{M} \ ,  
\end{equation}
with
\begin{equation}
  \label{eq:momentums}
  \bm{q} := m\bm{v} \ , 
  \quad
  \bm{h} := -\int_{P' \ \in \text{ body}} \hspace{-3em}   \bm{GP'}\times(\bm{GP'} \times \bm{\omega} ) \ dm = \bm{J}\! . \bm \omega \ , 
\end{equation} 
where $\bm{J}\!.$ denotes the inertia operator at $G$. Throughout this paper aircraft are modeled as rigid bodies of constant mass and 
we focus on the class of vehicles controlled via
four control inputs, namely the thrust intensity $T\in \mathbb{R}$ of a body-fixed thrust force $\bm{T}=-T\bm{k}$ and the three
components (in body-frame) of a control torque vector $\bm{\Gamma}_G$. This class of systems covers (modulo an adequate choice of control
inputs) a large variety of aerial vehicles of interest, like multi-copters, helicopters, convertibles UAVs, or even conventional airplanes.
The torque actuation can be obtained in various ways by using, e.g., 
control surfaces (fixed-wing aircraft), propellers (multi-copters), swash-plate mechanisms and tail-rotors (helicopters). 
By neglecting round-earth effects and buoyancy forces\footnote{The aircraft is assumed to be much heavier than air.}, 
control and external forces and moments acting on the aircraft are commonly modeled as follows 
\cite[Ch. 2]{fo94}, \cite{hhms09}, \cite{2004_STENGEL}, \cite{2003_STEVENS}:

\begin{equation}
\label{MeFeAircraft}
\begin{array}{lcl}
  \bm{F} &=& m \bm g +\bm{F}_{a} -T\bm{k} +\bm{F}_b\ ,  \\
  \bm{M} &=& \bm{GP}\times \bm{F}_a  +T\bm{k} \times \bm{G \Theta}+\bm{\Gamma}_G\ , 
\end{array}
\end{equation}
where $\bm g = g \bm k_0$ is the gravitational acceleration vector, $(\bm{F}_a,P)$ is the resultant of the 
aerodynamic forces and its point of application\footnote{The point $P$  is the so called 
\emph{body's center of pressure}. 
},
and $\Theta$ is the point of application of the thrust force.
In Eq.~\eqref{MeFeAircraft} we assume that the gyroscopic torque (usually associated with rotary-wing aircraft) is negligible or that it has already been compensated for via a preliminary torque control action. 
The force $\bm{F}_b$  is referred to as a \emph{body force}. It is induced by the control torque vector $\bm{\Gamma}_G$ and thus 
represents the effect of the control torque actuation on the position dynamics. The term $T\bm{k} \times \bm{G \Theta}$
in \eqref{MeFeAircraft} represents the effect of the control force actuation on the orientation dynamics. 

Beside the gravitational force, Eq. \eqref{MeFeAircraft} allows one to identify three types of forces (and torques):
{\it i)} {\em control forces},
{\it ii)} {\em body forces}, which cover coupling effects between thrust and torque actuations,
and {\it iii)} {\em aerodynamic forces}. This decomposition is based on a separation principle that is only valid in the first 
approximation. Nevertheless, identifying the dominant terms is useful 
from a control point of view to work out generic control strategies that can be refined on a 
case by case basis for specific classes of vehicles. A more detailed discussion of the modelling of body and aerodynamic forces follows.

\subsection{Body forces}

The influence of the torque control inputs on the translational dynamics via the body force 
$\bm{F}_b$ depends on the torque generation mechanism. 
More specifically, this coupling term is  negligible  for quadrotors \cite{hmlo02},
\cite{pmc10cep}, \cite{bms09ECC}, but it can be significant for helicopters 
due to the swashplate mechanism \cite[Ch.1]{hua09thesis}, \cite{dhl02}, \cite{ks98}, \cite{mhd99}, \cite[Ch. 5]{saber01},
and for ducted-fan tail-sitters due to the rudder system \cite[Ch. 3]{pf06}, \cite{psh04}. 
Thus, the relevance of this body force  must be discussed in relation to the specific application \cite{psh04} \cite[Ch. 3]{pf06} 
\cite{hhms13}. Note, however, that the body force $\bm{F}_b$ is typically small compared to either the gravitational force, the 
aerodynamic force, or the thrust force. Similarly, the term $T\bm{k} \times \bm{G \Theta}$ in \eqref{MeFeAircraft}, which reflects 
the influence of the thrust control input on the rotational dynamics, is usually small because $\Theta$ is close to the axis $(G,\bm k)$.
Assuming that body forces and corresponding torques can be either neglected or compensated for by control actions,
we focus hereafter on the modelling of aerodynamic forces acting on the vehicle's main body.

\subsection{Aerodynamic forces}

The modelling of aerodynamic forces and torques $\bm{F}_a$ and $\bm{M}_a := \bm{GP}\times \bm{F}_a$  acting on the vehicle 
is of particular importance.  Results on this topic can be found in \cite{2010_AND} 
\cite[Ch. 2]{2004_STENGEL} \cite[Ch. 2]{2003_STEVENS} for fixed-wing aircraft, in
\cite{hhwt09ICRA} \cite{bms09ECC} for quadrotors, in
\cite{jt06}  \cite{kog07} \cite{na08} \cite[Ch. 3]{pf06} \cite{pf10cep} for ducted-fan tail-sitters,  
and in \cite{2005_PROUTY}, \cite{vi03} for helicopters. 
The notation for aerodynamic forces used throughout this paper is presented next.

Denote by $\bm{v}_a$ the \emph{air velocity}, which is defined as the difference between $\bm{v}$ and 
wind's velocity $\bm{v}_w$, 
i.e. $\bm{v}_a = \bm{v} - \bm{v}_w$.
The \emph{lift force} $\bm{F}_L$ is the aerodynamic force component perpendicular to the air velocity,
and the \emph{drag force} $\bm{F}_D$ is the aerodynamic force component along the air velocity's direction. Now,
consider a (any) pair of angles $(\alpha,\beta)$ characterizing the orientation of $\bm{v}_a$ with respect to the body frame 
(e.g. Figure~\ref{fig:aerodyEff}). Combining the \emph{Buckingham $\pi-$theorem}~\cite[p. 34]{2010_AND} with the 
knowledge that the intensity of the \emph{steady} 
aerodynamic force varies approximately like the square of the air speed $|\bm{v}_a|$  yields the existence of two dimensionless 
functions ${C}_L(\cdot)$ and ${C}_D(\cdot)$ depending on the \emph{Reynolds number}
$R_e$, the \emph{Mach number} \emph{M}, and $(\alpha,\beta)$, and such that
\begin{equation}
\label{eq:FaComponents}
\begin{array}{l}
\bm{F}_a  =  \bm{F}_L +\bm{F}_D,   \\ 
\bm{F}_L  =  k_a|\bm{v}_a|C_L(R_e,M,\alpha,\beta)\bm{r}(\alpha,\beta,|\bm v_a|) \times \bm{v}_a,  \\
\bm{F}_D  =  -k_a |\bm{v}_a|C_D(R_e,M,\alpha,\beta)\bm{v}_a,  \\
\bm{r} \cdot \bm{v}_a  =  0,  \quad |\bm{r}| = 1,  \\
k_a := \rho \Sigma/2, 	  
\end{array}
\end{equation}
with $\rho$ the \emph{free stream} air density, $\Sigma$ an area germane to the given body shape,
$\bm{r}(\cdot)$ a unit vector-valued function, $C_D$ ($\in \mathbb{R}^+$) and  $C_L$ ($\in \mathbb{R}$) 
the \emph{aerodynamic characteristics} of the body, i.e. the so-called \emph{drag coefficient} and \emph{lift coefficient}, respectively. 
In view of the above representation of the aerodynamic force -- first introduced in~\cite{phms12} -- 
the lift direction is independent from the aerodynamic coefficients, which in turn characterize the aerodynamic force intensity
since $|\bm{F}_a| = k_a |\bm{v}_a|^2\sqrt{C^2_L + C^2_D}$. The lift direction is fully characterized by the unitary 
vector $\bm{r}(\cdot)$, which only depends on $(\alpha, \beta)$ and on the air velocity magnitude $|\bm v_a|$. 
We will see further on that axisymmetry of the vehicle's body yields a specific expression of the vector $\bm{r}(\cdot)$.
By considering the model~\eqref{eq:FaComponents}, we implicitly neglect
the effects of the vehicle's \emph{rotational and unsteady motions} on its surrounding airflow (see~\cite[p. 199]{2004_STENGEL} for more details). 

\subsection{Control model}
With the assumptions and simplifications discussed above, the control model reduces to
\begin{IEEEeqnarray}{RCL}
  \label{eq:newton0-1} 
  m \dot{ \bm v} &=& m\bm{g} +\bm{F}_a -T\bm{k}, \IEEEyessubnumber  \\
  \frac{d}{dt} (\bm{i},\bm{j},\bm{k}) &=& \bm{\omega} \times (\bm{i},\bm{j},\bm{k}), \IEEEyessubnumber \\ 
  \frac{d}{dt} (\bm{J} \! . \bm{\omega}) &=& \bm{GP} \times \bm F_a + \bm{\Gamma}_G \label{eq:newton0-3}.
\end{IEEEeqnarray}

To develop general control principles that apply to a large number of aerial vehicles, 
one must get free of actuation specificities and concentrate on the vehicle's governing dynamics. 
In agreement with a large number of works on VTOL control (see~\cite{hhms13} for a survey)
and in view of Eq.~\eqref{eq:newton0-3}, which points out how $\bm \omega$ can be modified via the choice
of the control torque $\bm{\Gamma}_G$, a complementary assumption consists in considering 
the angular velocity $\bm{\omega}$ as an intermediate control input.
This implicitly means that the control torque calculation and production can be done independently of high-level control 
objectives, at least in the first design stage. 
The corresponding physical assumption is that ``almost'' any desired angular velocity can 
be obtained after a short transient time. 
In the language of Automatic Control, this is a 
typical ``backstepping'' assumption. 
Once it is made, the vehicle's actuation consists in four input variables, 
namely, the thrust intensity and the three components of $\bm{\omega}$. 
The control model then reduces to Eqs.~\eqref{eq:newton0-1}, 
with $T$ and $\bm{\omega}$ as control inputs. 

\section{Symmetric bodies and spherical equivalence}
\label{sec:spheEquiv}

Eq. \eqref{eq:newton0-1} shows how the gravitational force $m \bm g$ and  the aerodynamic force $\bm{F}_a$ 
take part in the body's linear acceleration vector. It also shows that, for the body to move with a constant velocity,
 the controlled thrust vector $T \bm{k}$ must be equal to the resultant external force 
\[\bm{F}_{ext} :=m \bm{g}+\bm{F}_a.\]
When $\bm{F}_a$ does not depend on the vehicle's orientation, as in the case of \emph{spherical bodies} (see \cite{hua09thesis} for details), 
the resultant external force $\bm{F}_{ext}$ does not depend on this orientation either. 
The thrust direction at the equilibrium is then unique 
and it is explicitly given by the direction of $\bm{F}_{ext}$.
The control strategy then basically consists in aligning  the thrust direction $\bm{k}$ with the direction of $\bm{F}_{ext}$ 
(using $\bm \omega$ as control input) and in opposing the thrust magnitude to the intensity of $\bm{F}_{ext}$ (using the thrust $T$ as control input). 
This is the basic principle of the thrust direction control paradigm \cite{ghm05,hhms09}. For most vehicles encountered in practice, however, 
aerodynamic forces depend on the vehicle's orientation, and thus on the direction of $\bm k$.  In particular, the equilibrium relation 
$T \bm k = \bm{F}_{ext} $
then becomes an implicit equation with both sides of this equality depending on $\bm k$. 
In this case, existence, uniqueness, and explicit determination of the equilibrium thrust direction(s) become fundamental 
questions for the control design \cite{pucciPhd}. In this section, we provide answers to these 
questions for a class of axisymmetric vehicles, in the continuity of  \cite{pucciPhd}, \cite{phms12}, where axisymmetry
is shown to infer geometrical aerodynamic properties that simplify the associated control problem.
More precisely, let us consider vehicles whose external surface $\mathcal{S}$ is characterized by the existence of
an orthonormal body frame $\mathcal{B}_c = \{G_c;\bm{i},\bm{j},\bm{k}\}$
that satisfies either one of the following assumptions:

\begin{figure}[t!]
\centering
\begin{minipage}[t]{0.47\linewidth} 
    \centering
    \def\svgwidth{1.05\linewidth}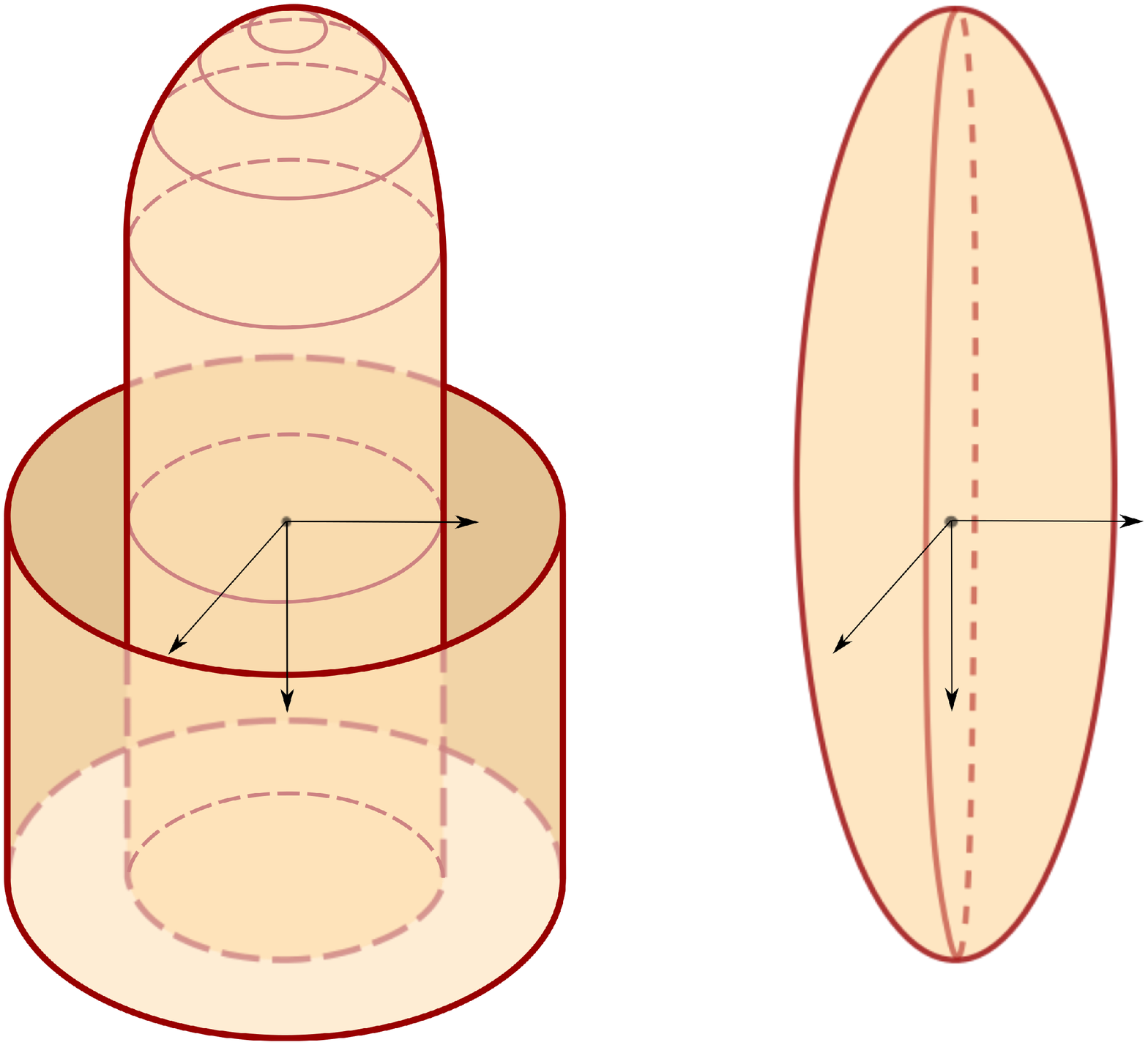
    \caption{Symmetric and bisymmetric shapes}\label{fig:symmetric}
\end{minipage}
\quad
\begin{minipage}[t]{0.4\linewidth} 
    \centering
    \def\svgwidth{.7\linewidth}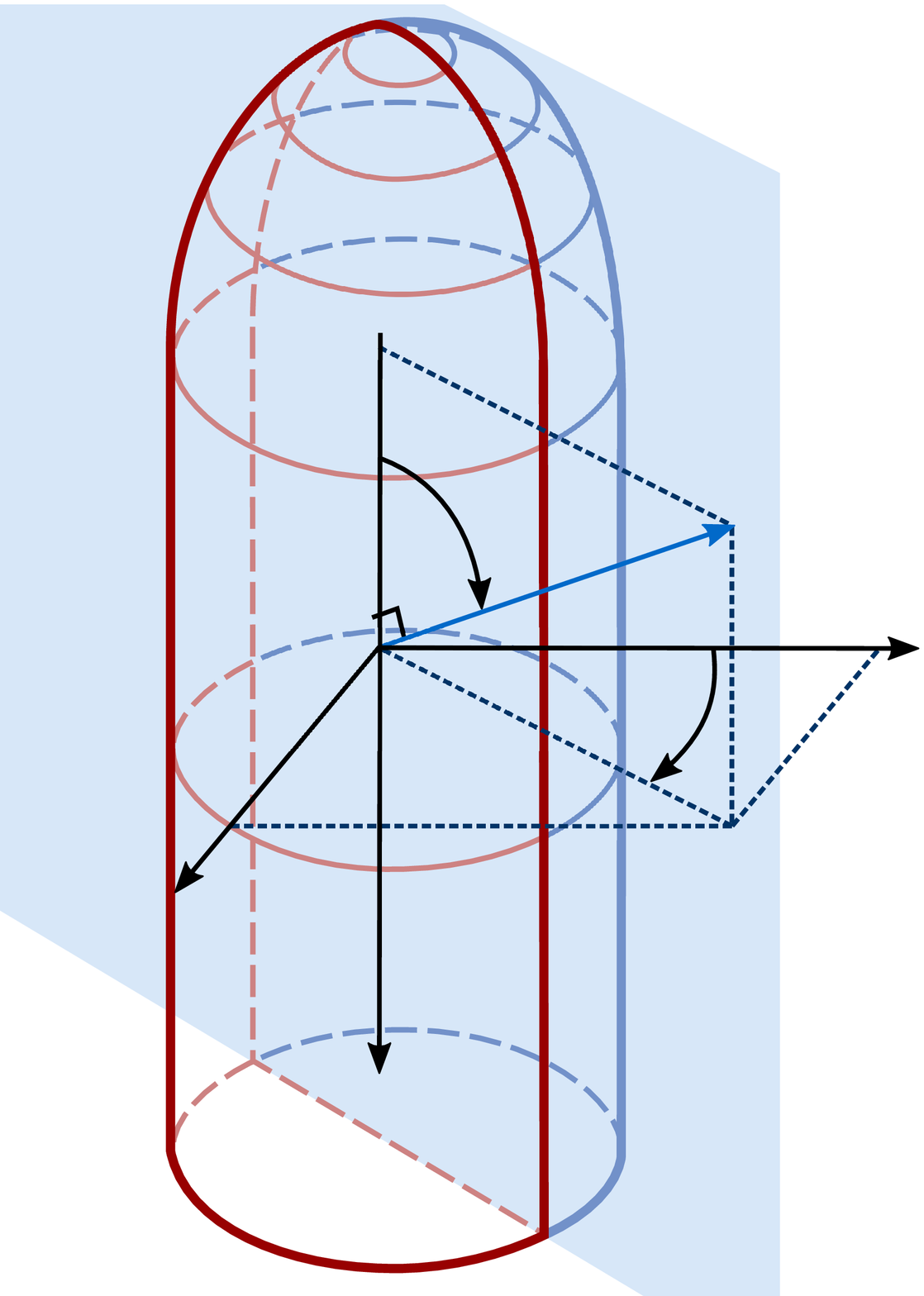
    \caption{The $(\alpha,\beta)$ angles.}
    \label{fig:aerodyEff}
\end{minipage}
\end{figure}

\begin{hypothesis}[\textbf{Symmetry}]
  \label{hy:symmetries}
  Any point $P \in \mathcal{S}$ transformed by the rotation of an angle $\theta$ about the axis $G_c\bm{k}$, i.e. by the operator defined by
 $g_\theta(\cdot) = \rot_{G_c\bm{k}}(\theta)(\cdot)$, also belongs to $\mathcal{S}$, i.e. $g_\theta(P) \in \mathcal{S}$.
\end{hypothesis}
\begin{hypothesis}[\textbf{Bisymmetry}]
  \label{hy:bisymmetries}
  Any point $P \in \mathcal{S}$ transformed by the composition of two rotations of angles $\theta$ and $\pi$ about the axes 
$G_c\bm{k}$ and $G_c\bm{j}$, i.e. by the operator defined by 
$g_\theta(\cdot) = (\rot_{G_c\bm{k}}(\theta) \circ \rot_{G_c\bm{\jmath}}(\pi))(\cdot)$, 
  also belongs to $\mathcal{S}$, i.e. $g_\theta(P) \in \mathcal{S}$.
\end{hypothesis}
The operator $\rot_{O\bm{v}}(\psi)(P)$ stands for the rotation about the axis $O\bm{v}$ by the angle $\psi$ of the point $P$.
Examples of ``symmetric'' and ``bisymmetric'' shapes satisfying these assumptions are represented in Figure~\ref{fig:symmetric} (with $G=G_c$).
Note that various human-made aerial devices (rockets, missiles, airplanes with annular wings, etc.) satisfy the symmetry property of Assumption  
\ref{hy:symmetries} in the first approximation, and that the present study is thus of direct relevance for the modelling and control of these devices.
For symmetric shapes, i.e. such that Assumption~\ref{hy:symmetries} holds true, one can define $\alpha \in [0,\pi]$ as the 
{\em angle of attack}\footnote{The angle of attack $\alpha$ so defined does not coincide with that used for airplanes equipped with planar 
wings, which break the body's rotational symmetry about $G_c\bm{k}$~\cite[p. 53]{2004_STENGEL}.} between $-\bm{k}$ and $\bm{v}_a$, and 
$\beta \in (-\pi,\pi]$ as the angle between the unit frame vector $\bm{i}$ and the projection of $\bm{v}_a$ on the plane 
$\{G_c;\bm{i},\bm{j}\}$ (see Figure~\ref{fig:aerodyEff}). Observe that this assumption also implies that:

\noindent
$\mathbf{P1:}$ the aerodynamic force $\bm{F}_a$ does not change when the body rotates about its axis of symmetry $G_c\bm{k}$;

\noindent
$\mathbf{P2:}$ $\bm{F}_a \in \text{span}\{\bm{k},\bm{v}_a\}$ .

Property P1 in turn implies that the aerodynamic characteristics do not depend on $\beta$, whereas Property P2 implies that 

$i)$ the unit vector $\bm{r} \ $ in~\eqref{eq:FaComponents}  is orthogonal to $\bm{k}$ and independent of the angle of attack $\alpha$; 


$ii)$ the lift coefficient is equal to zero when $\alpha = \{0,\pi\}$. 

Subsequently, the expressions~\eqref{eq:FaComponents} of the lift and drag forces specialize to
\begin{equation}
\label{eq:aerodyModelSymmetricBodies}
\begin{array}{lcl}
	\bm{F}_L & = & k_a|\bm{v}_a|C_L(R_e,M,\alpha)\bm{r}(\beta) \times \bm{v}_a, \\
	\bm{F}_D & = & -k_a |\bm{v}_a|C_D(R_e,M,\alpha)\bm{v}_a, \\
        \bm{r}(\beta) & = & -\sin(\beta) \bm{i}+\cos(\beta) \bm{j}.
\end{array}
\end{equation}
Under the stronger Assumption~\ref{hy:bisymmetries}, i.e. when the body's shape is also $\pi$-symmetric w.r.t. the $G_c\bm{\jmath}$ axis, 
the aerodynamic characteristics $C_L$ and $C_D$ must be $\pi-$periodic w.r.t. $\alpha$.
The aforementioned choice of $(\alpha,\beta)$ implies that 
\begin{equation}
\label{system:alphaBeta}
\alpha = \cos^{-1}\left(-\frac{v_{a_3}}{|\bm{v}_a|}\right),\quad
\beta = \mathrm{atan2}(v_{a_2},v_{a_1}),
\end{equation}
and
\begin{equation}
\label{system:vaDecomposition}
\begin{array}{lcl}
v_{a_1} &=& |\bm{v}_a|\sin(\alpha)\cos(\beta), \\
v_{a_2} &=& |\bm{v}_a|\sin(\alpha)\sin(\beta), \\
v_{a_3} &=& -|\bm{v}_a|\cos(\alpha).
\end{array}
\end{equation}
with $v_{a_i}$ ($i=1,2,3$) denoting the coordinates of $\bm{v}_a$ in the body-fixed frame basis.        
From the definitions of $\alpha$ and $\bm{r}(\beta)$, one then verifies that
\[\bm{r}(\beta) \times \bm{v}_a = -\cot(\alpha) \bm{v}_a-\frac{|\bm{v}_a|}{\sin(\alpha)}\bm{k},
\]
so that $\bm{F}_a = \bm{F}_L+\bm{F}_D$ becomes
\begin{equation}
\label{FaNewExpressSymm}
\bm{F}_a = {-}k_a |\bm{v}_a|\hspace{-0.1cm}\left[\hspace{-0.1cm} 
\Big(\hspace{-0.05cm}C_D(\cdot)+C_L(\cdot)\cot(\alpha)\Big)\bm{v}_a+\frac{C_L(\cdot)}{\sin(\alpha)}|\bm{v}_a| \bm{k} \right]. 
\end{equation}
For constant Reynolds and Mach numbers the aerodynamic coefficients depend only on $\alpha$ and one readily deduces the following result
from \eqref{FaNewExpressSymm}. 

\begin{proposition}[\cite{phms12}, \cite{pucciPhd}]
\label{th:conditionA}
Consider an axisymmetric thrust-propelled vehicle subjected to aerodynamic forces given by~\eqref{eq:aerodyModelSymmetricBodies}. 
Assume that the aerodynamic coefficients satisfy the following relation
\begin{equation}
		\label{eq:conditionOnalpha}
		C_D(\alpha)+ C_L(\alpha)\cot(\alpha) = C_{D_0},  
\end{equation}
with $C_{D_0}$ denoting a constant number.
Then, Eq. \eqref{eq:newton0-1} can also be written as
\begin{equation}
m\dot{\bm v}=m\bm{g}+ \bm{F}_p -T_p\bm{k}, 
\label{eq:newFormDynamics} 
\end{equation}
with

\begin{IEEEeqnarray}{RCL}
\label{eq:fpGen}
  T_p &=& T + k_a|\bm{v}_a|^2\frac{C_L(\alpha)}{\sin(\alpha)}, \IEEEyessubnumber \\ 
  \bm{F}_p(\bm v_a) &=& -k_a C_{D_0}|\bm{v}_a| \bm{v}_a.   \IEEEyessubnumber
\end{IEEEeqnarray}

\end{proposition}
This proposition points out the possibility of seeing an axisymmetric body subjected to both drag and lift forces as a 
sphere subjected to the orientation independent drag force $\bm{F}_p$ and powered by the thrust force $\bm{T}_p=-T_p\bm{k}$. 

It follows from 
\eqref{eq:newFormDynamics} that  given a desired reference velocity $\bm v_r$, there exists a unique (up to sign) equilibrium 
thrust direction $\bm k_{ref}$ as long as $|m\bm{g}+ \bm{F}_p -m \dot{\bm v}_r | \ne 0 $ along this reference velocity. In particular,  
this direction is explicitly defined by 
\[\bm k_{ref}= \frac{m\bm{g}+ \bm{F}_p(\bm v_{r,a})-m \dot{\bm v}_r}{|m\bm{g}+ \bm{F}_p(\bm v_{r,a})-m \dot{\bm v}_r|}\] where $\bm v_{r,a}= \bm v_r -\bm v_w$.
The main condition for this result to hold is that the relation~\eqref{eq:conditionOnalpha} must be satisfied. 
Obviously, this condition is compatible with an infinite number of functions $C_D$ and $C_L$. 
Let us 
point out  a particular set of simple functions
that also satisfy the $\pi$-periodicity property w.r.t. the angle of attack $\alpha$ associated with \emph{bisymmetric}  bodies. 
\begin{proposition}
\label{prop:2D->3D}
The functions $C_D$ and $C_L$ defined by 
\begin{IEEEeqnarray}{RCL}
\label{system:aerodynamicsChaPart}
  C_D(\alpha) &=& c_0 + 2c_1\sin^2(\alpha) \IEEEyessubnumber \\
C_L(\alpha) &=&  c_1\sin(2\alpha), \IEEEyessubnumber
\end{IEEEeqnarray}
with $c_0$ and $c_1$ 
two real numbers, satisfy the condition \eqref{eq:conditionOnalpha} with 
$C_{D_0} = c_0 + 2c_1.$ 
The equivalent drag force and thrust intensity are then given by
\begin{IEEEeqnarray}{RCL}
\label{eq:parapetersNFD}
\bm F_p(\bm v_a) &=& -k_a C_{D_0} |\bm{v}_a| \bm{v}_a, \IEEEyessubnumber  \\ 
T_p &=& T + 2c_1k_a|\bm{v}_a|^2\cos(\alpha).  \IEEEyessubnumber
\end{IEEEeqnarray}
\end{proposition}

\begin{figure}[t]
 \centering
 \small{\input{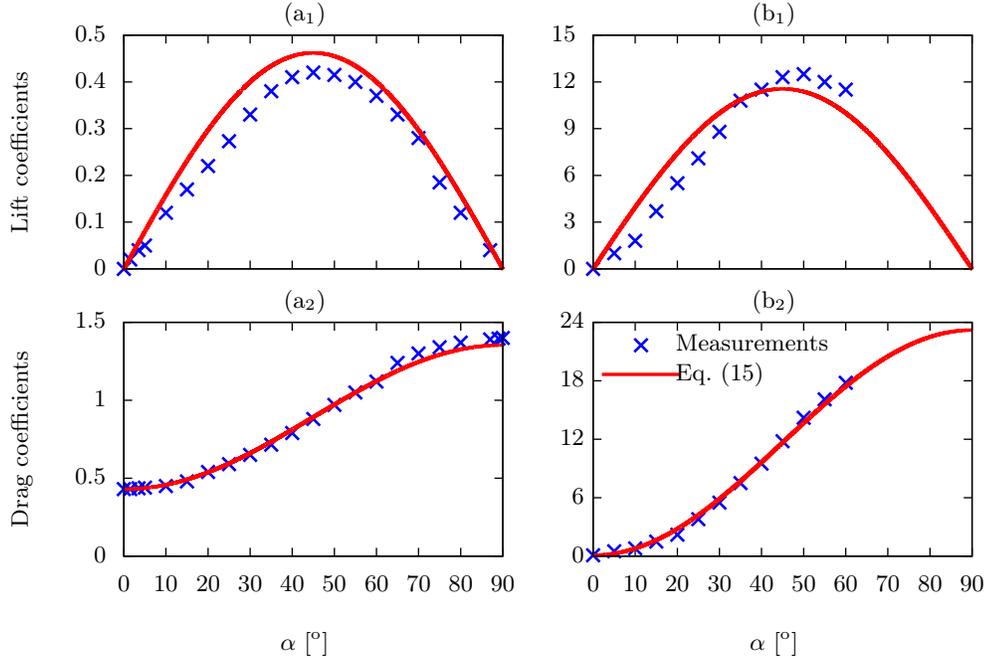}}
 \caption{Aerodynamic coefficients of:  (a$_{1,2}$) elliptic bodies; (b$_{1,2}$) missile-like bodies.}
 \label{fig:ellipticAndMissile}
\end{figure}

The proof is straightforward. A particular bisymmetric body is the sphere whose aerodynamic characteristics (zero lift coefficient and constant drag coefficient) are obtained 
by setting $c_1=0$ in \eqref{system:aerodynamicsChaPart}. Elliptic-shaped bodies are also bisymmetric but, in contrast with the sphere, 
they do generate lift in addition to drag. The process of approximating measured aerodynamic characteristics with 
functions given by \eqref{system:aerodynamicsChaPart} is illustrated by the Figure~\ref{fig:ellipticAndMissile}(a) where we have used 
experimental data borrowed from \cite[p.19]{1965_WAYNE} for an elliptic-shaped body with Mach and Reynolds numbers equal to $M = 6$ 
and $R_e = 7.96\cdot 10^6$ respectively. For this example, the identified coefficients are $c_0=0.43$ and $c_1=0.462$. Since missile-like 
devices are ``almost'' bisymmetric, approximating their aerodynamic coefficients with such functions can also be attempted. For instance, 
the approximation shown in Figure~\ref{fig:ellipticAndMissile}(b) has been obtained by using experimental data taken from \cite[p.54]{1971_SAFFEL} 
for a missile moving at $M = 0.7$. In this case, the identified coefficients are $c_0=0.1$ and $c_1=11.55$. In both cases, the match between 
experimental data and the approximating functions, although far from perfect, should be sufficient for feedback control purposes.

Note that the process of approximating aerodynamics characteristics by trigonometric functions is not new (see, e.g., \cite{dls99,wa08}).
To our knowledge, however, such approximations have not been exploited for the explicit determination of equilibrium orientations,
as deduced from Proposition \ref{th:conditionA}.

\section{Control design}
\label{sec-control}
The results of the previous section are now exploited to address feedback control design of axisymmetric vehicles.
We first start by considering the thrust direction control problem. Several solutions to this problem have already been proposed in 
the literature. The solution proposed hereafter is a coordinate-free extension of the solution given in \cite{hhms09}.

\subsection{Thrust direction control}

Consider a time-varying reference thrust (unitary) direction $\bm k_r$. It is assumed that $\bm k_r$ varies 
smoothly with time so that $\dot{\bm k}_r(t)$ is well defined for any time $t$. The following result provides control expressions 
for the angular velocity control input $\bm \omega$ yielding a large stability domain.

\begin{proposition}
\label{prop-thrust-cont}
The feedback law 
\begin{equation}
\label{def-omega}
\bm \omega = \left( k_1(\bm k, t) + \frac{\dot \gamma(t)}{\gamma(t)} \right) \bm k \times \bm k_r +\bm \omega_r + \lambda(\bm k, t) \bm k
\end{equation}
with $\bm \omega_r = \bm k_r \times \dot{\bm k}_r$, $\lambda(\cdot)$ any real-valued continuous function, 
$\gamma(\cdot)$ any smooth positive real-valued function such that $\inf_t \gamma(t) >0$,
and $k_1(\cdot)$ any continuous positive real-valued function such that $\inf_{\bm k,t} k_1(\bm k,t) >0$,
ensures exponential stability
of the equilibrium $\bm k= \bm k_r$ with domain of attraction $\{\bm k(0): \bm k(0) \cdot \bm k_r(0) \neq -1 \}$.
\end{proposition}
The proof is given in the appendix.
\vspace{.3cm}

\noindent The above expression of $\bm \omega$ is a generalization of the solution proposed in \cite{hhms09}, for which the control 
gain $\gamma$ was not present and a specific choice of $k_1$ was imposed. The additional degrees of freedom given by the above solution
will be exploited further on.
Recall that the limitation on the stability domain is due to the topology of the unit sphere, which forbids the existence of 
smooth autonomous feedback controllers yielding global asymptotic stability. The first term in the right-hand side of \eqref{def-omega} 
is a nonlinear feedback term that depends on the
error between $\bm k$ and $\bm k_r$, here given by the cross product of these two vectors. The second term is a feedforward term. In practice, 
this term can be neglected when the vector $\dot{\bm k}_r$ (and thus $\bm \omega_r$) is not known, as in the case where $\bm k_r$ 
corresponds to a reference thrust direction manually specified by a human pilot using a joystick. 
Omitting this feedforward term is not very damaging in terms of performance, provided that $\bm k_r$ does not vary too rapidly. 
Finally, the last term in the right-hand side of \eqref{def-omega} is associated with the rotation about the axis 
$\bm k$ (yaw degree of freedom for a hovering VTOL vehicle, and roll degree of freedom for a missile or for a cruising airplane with annular wing). It does not affect the thrust direction dynamics since $\dot{\bm k}= \bm \omega \times \bm k$. 
Finally, let us comment on the choice of the control gains. Concerning $\lambda(\cdot)$, the simplest choice is obviously
$\lambda(t) \equiv 0$. Another possibility is $\lambda(t)= - \bm \omega_r(t) \cdot \bm k(t)$. This yields 
$\bm \omega(t) \cdot \bm k(t) =0 \; \forall t$ so that the control law does not induce any instantaneous rotation around $\bm k$. 
Other choices may be preferred when it matters to precisely control the vehicle's remaining rotational degree of freedom.
Concerning $\gamma$ and $k_1$, a simple choice consists in taking constant positive numbers, but other possibilities can be preferable. 
For instance, taking $k_1(\bm k,t)=k_{1,0}/(1+\bm k \cdot \bm k_r(t)+\epsilon_1)$, with $k_{1,0}>0$ and $\epsilon_1$ a small positive number, makes the feedback gain 
$k_1$ grow large when $\bm k$ gets close to $- \bm k_r$ and, subsequently, tends to make this undesired equilibrium direction more repulsive. As for 
$\gamma$, a choice adapted to the objective of tracking reference trajectories, in either position or velocity, is pointed out thereafter. 

\subsection{Velocity and position control for axisymmetric vehicles}
\label{sec:velocitycontrol}

In what follows, $\bm{v}_r(\cdot)$ denotes a reference velocity time-function (at least three times differentiable everywhere). 
Velocity control then consists in the asymptotic stabilization of the velocity error $\bm{\tilde{v}}:=\bm{v}-\bm{v}_r$ at zero. 
This control objective may be complemented by the convergence to zero of a position error $\bm{\tilde{p}}:=\bm{p}-\bm{p}_r$, with $\bm{p}_r(\cdot)$ 
denoting a reference position time-function. In this latter case,  $\bm{v}_r$ is the time-derivative of $\bm{p}_r$, and the error state vector to be 
stabilized at zero contains the six-dimensional vector $(\bm{\tilde{p}},\bm{\tilde{v}})$. The error vector may further include an integral of the position 
error $\bm{\tilde{p}}$. It is also possible that the application only requires the stabilization of the vehicle's altitude, in addition to its velocity. 
In order to take various control objectives involving the vehicle's velocity and possibly other state variables whose variations depends on this velocity, 
we consider from now on a ``generalized'' control objective consisting in the asymptotic stabilization at zero of an error vector denoted as $(\bm{\tilde{\rho}},\bm{\tilde{v}})$, 
with $\bm {\tilde \rho} \in \RR^p$ and such that $\dot{\bm {\tilde \rho}} = \bm f(\bm {\tilde \rho}, \bm {\tilde v})$, with $\bm f(\cdot,\cdot)$ denoting a smooth vector-valued 
function. 
For instance, in the case where $\bm{\tilde{\rho}}= \bm{\tilde{p}}$, with $\bm{\tilde{p}}$ denoting either a position error, or an integral of the velocity error $\bm {\tilde v}$,  
then $\bm f(\bm {\tilde \rho}, \bm {\tilde v})=\bm {\tilde v}$. If $\tilde{\bm \rho} = (\bm I_p,\tilde{\bm p})$, with $\bm I_p$ denoting
a saturated integral of the position tracking error such that $\frac{d}{dt} \bm I_p = h(\tilde{\bm \rho})$, then  
$\bm f(\bm {\tilde \rho}, \bm {\tilde v})=(h(\tilde{\bm \rho}),\bm {\tilde v})$. The simplest case corresponds to pure velocity control without integral correction, 
for which $\tilde{\bm \rho} = \emptyset$.

Consider now an axisymmetric vehicle with its velocity dynamics given by \eqref{eq:newFormDynamics}, and let $\bm{a}_r:=\dot{\bm v}_r$ denote the reference acceleration. 
It follows from \eqref{eq:newFormDynamics} that
\begin{equation}
\label{def-dv}
m \dot{\bm {\tilde v}}= \bm F_p + m(\bm{g} - \bm a_r) -T_p\bm{k}.
\end{equation}
Introducing an auxiliary feedback term $\bm{\xi}$, whose role and choice will be commented upon thereafter, this equation can be written as 
\begin{equation}
\label{def-mdvt}
m \dot{\bm {\tilde v}} = m \bm{\xi} + \bar{\bm F}_p -T_p\bm{k},
\end{equation}
with
\begin{equation} \label{Force}
\bar{\bm F}_p(\bm v_a,\bm a_r,\bm{\xi}):=\bm{F}_p(\bm v_a)+m(\bm{g}-\bm{a}_r -\bm{\xi})
\end{equation}
The idea is to end up working with the simple control system $\dot{\bm {\tilde v}} = \bm{\xi}$. To this aim
Eq.~\eqref{def-mdvt} suggests to adopt a control strategy that ensures
the convergence of $\bar{\bm F}_p - T_p \bm k$ to zero. With $T_p$ preferred positive, this implies that the thrust direction $\bm k$ should tend to
\begin{equation}
\label{def-kr}
\bm k_r:=\frac{\bar{\bm F}_p}{|\bar{\bm F}_p|}~.
\end{equation}
Recall from \eqref{eq:fpGen} that $\bm F_p$ does not depend on $\bm k$. Thus, provided that $\bm{\xi}$ does not depend on $\bm k$, 
$\bar{\bm F}_p$ does not depend on $\bm k$ either, and $\bm k_r$ is well defined as long as $\bar{\bm F}_p$ does not vanish. 
This is precisely what makes Proposition \ref{th:conditionA} important for the control design. 
Convergence of  $\bar{\bm F}_p - T_p \bm k$ to zero also implies that $T_p$ must tend to $\bar{\bm F}_p \cdot \bm k$. 
From \eqref{eq:newton0-1} and \eqref{eq:newFormDynamics}, this is equivalent to the convergence of the thrust intensity $T$ to $\bar{\bm F}_a \cdot \bm k$
with
\begin{equation}
\label{def-fab}
\bar{\bm F}_a:= \bm{F}_a+m(\bm{g}-\bm{a}_r -\bm{\xi})
\end{equation}
Once the reference thrust direction $\bm k_r$ is properly defined, a possible control law, among other possibilities, is pointed out in the following proposition.

\begin{proposition}
\label{prop-velocity-cont}
Consider an axisymmetric vehicle for which the aerodynamic characteristics satisfy relation \eqref{eq:conditionOnalpha}, and a smooth feedback controller $\bm{\xi}(\bm {\tilde \rho}, \bm {\tilde v})$ for the control system 
\begin{subequations}
\label{def-fullyactuated}
\begin{align}
\dot{\bm {\tilde \rho}} & =  \bm f(\bm {\tilde \rho}, \bm {\tilde v}) \label{def-fullyactuated-1}\\
\dot{\bm {\tilde v}} & =  \bm{\xi}
\end{align}
\end{subequations}

Assume that 
\begin{description}
\item [A1~:] $\bm{\xi}(\bm {\tilde \rho}, \bm {\tilde v})$ makes 
$(\bm {\tilde \rho},\bm{\tilde v})=(\bm 0, \bm 0)$ a locally exponentially stable equilibrium point 
of System \eqref{def-fullyactuated};
\item [A2~:] $\bar{\bm F}_p$ does not vanish along the velocity reference trajectory $\bm v_r$, i.e., $\exists \delta>0~:~\delta \leq \bar{\bm F}_p(\bm v_{r,a}(t),\bm a_r(t), \bm 0), \; \forall t$, with $\bm v_{r,a}:= \bm v_r-\bm v_w$.
\end{description}
Then, $T= \bar{\bm F}_a \cdot \bm k$ and $\bm \omega$ given by \eqref{def-omega}, with
$\bm k_r$ defined by \eqref{def-kr},
$\gamma = \sqrt{c_2+ |\bar{\bm F}_p|^2}$, and $c_2$ any strictly positive constant, ensure
local exponential stability of the equilibrium point $(\bm{\tilde \rho}, \bm v, \bm k)= (\bm 0, \bm v_r, \bm k_r)$
for the system \eqref{def-fullyactuated-1}-\eqref{eq:newton0-1}.
\end{proposition}
The proof is given in the appendix.
\vspace{.3cm}

\noindent Let us comment on the  above result.
\begin{enumerate}
\item Proposition \ref{prop-velocity-cont} essentially shows how to derive an exponentially stabilizing feedback law for the underactuated
System \eqref{eq:newton0-1} from an exponentially stabilizing feedback controller for 
the fully-actuated system $\dot{\bm {\tilde v}}= \bm \xi$. 
Since feedback control of fully-actuated systems can be addressed
with a large variety of existing control laws, starting with linear feedback control, the determination of $\bm \xi$ will not be further addressed here. 
\item Once an exponential stabilizer $\bm \xi$ of the origin of System \eqref{def-fullyactuated} is determined, local exponential stability of zero tracking errors for an antisymmetric vehicle for which the aerodynamic characteristics satisfy relation \eqref{eq:conditionOnalpha} essentially 
relies on Assumption 2, which imposes that the reference thrust direction $\bm k_{ref}$, associated with perfect tracking of the reference trajectory, is well defined at all times. This condition may be violated
for very specific and aggressive reference trajectories. Note, however, that its satisfaction can be checked from the knowledge of the reference velocity
only (assuming of course that an accurate model of aerodynamic forces is available). 
\item Finally, let us discuss a few issues related to the calculation of the feedback control. The main difficulty at this level comes 
from the fact that both $\bm k_r$ and $\gamma$ depend on $\bar{\bm F}_p$.
Since $\dot \gamma$ and $\dot{\bm k}_r$ are involved in the calculation of $\bm \omega$, the time-derivative of $\bar{\bm F}_p$ has to be calculated also.
In practice, a possibility consists in estimating this term, e.g. from 
the calculation of $\bar{\bm F}_p$ and using a high-gain observer. Another possible choice, consisting in using the reference velocity instead of the vehicle's actual velocity to calculate an approximation of this term, is made for the simulations reported in the next section.
\end{enumerate}

Proposition \ref{prop-velocity-cont} guarantees {\em local} asymptotic stability only. The difficulty to ensure a large domain of convergence comes from the risk of $\bar{\bm F}_p$ vanishing at some point, which would in turn make $\bm k_r$, as specified by \eqref{def-kr}, 
ill-defined. This risk, although small, cannot be ruled out in the most general situation, especially because the term $\bm F_p$  in $\bar{\bm F}_p$ (i.e. the term resulting from the aerodynamic forces acting on the vehicle) can take very large values. 
In practice, the necessity of having a control always well defined implies that one has to modify the term $\bar{\bm F}_p$ used in the control expression in order to avoid its passage through zero. A reasonable way of making this modification is a subject of future studies. 
Taking the above-mentioned difficulty aside, if one assumes that $\bar{\bm F}_p$ remains different from zero, then convergence
of the tracking errors can be guaranteed, as specified by the following proposition.

\begin{proposition}
Given the feedback law of Proposition \ref{prop-velocity-cont}, if one further assumes that
\begin{description}
\item [A1(bis)~:]~~ $\bm{\xi}(\bm {\tilde \rho}, \bm {\tilde v})$ globally asymptotically stabilizes 
the origin $(\bm {\tilde \rho},\bm{\tilde v})=(\bm 0, \bm 0)$ of the system
\[
\begin{array}{lcl}
\dot{\bm {\tilde \rho}} & =  & \bm f(\bm {\tilde \rho}, \bm {\tilde v}) \\
\dot{\bm {\tilde v}} & = & \bm{\xi}(\bm {\tilde \rho}, \bm {\tilde v}) + \varepsilon(t)
\end{array}
\]
when the "perturbation" $\varepsilon(\cdot)$ is identically zero, and still ensures the convergence to zero of the solutions to this system 
when $\varepsilon(\cdot)$ converges to zero exponentially;
\end{description}
then any solution to the closed-loop system 
\eqref{def-fullyactuated-1}-\eqref{eq:newton0-1} 
along which $\bar{\bm F}_p$ does not vanish (in the sense that $\exists \delta>0~:~\delta \leq \bar{\bm F}_p(\bm v_a(t),\bm a_r(t), \bm{\xi}(\bm {\tilde \rho}, \bm {\tilde v})), \; \forall t$) converges to 
the equilibrium point $(\bm 0, \bm v_r, \bm k_r)$.
\end{proposition} 
The proof follows directly from the proof of Proposition \ref{prop-velocity-cont}. Preservation of the convergence to zero of the system's solutions in the case of an exponentially decaying additive perturbation, although needed for the sake of completeness, is a weak requirement that has little impact on the control design. 

\subsection{Simulation results} \label{simulation}

The feedback law of Proposition \ref{prop-velocity-cont} is applied to a model of the C-701 anti-ship missile, 
whose geometry and operational characteristics are close to those of the device associated with the measured aerodynamic coefficients 
of Figure~\ref{fig:ellipticAndMissile}(b). The control objective is the asymptotic stabilization at zero of the velocity error $\tilde{\bm v}$.  A saturated integral  
$\bm{ \tilde \rho}=\bm{I}_v$ of this error is used in the control law in order to compensate for static modelling errors and additive perturbations. 
This integral term is obtained as the (numerical) solution to the following equation 
\cite{2011_HUA-SAMSON}~\cite{2005_SESHAGIRI-KHALIL}
\begin{equation} \label{eq: Iv}
\frac{d}{dt} \bm{I}_{v}= \bm f(\bm {\tilde{\rho}}, \bm {\tilde v})  =  -k_I \bm{I}_v + k_I \text{sat}^{\delta}\left(\bm{I}_v+\frac{ \bm{\tilde{v}}}{k_I}\right)~~;~\bm{I}_v(0)=0,
\end{equation}
with $k_I$ a (not necessarily constant) positive number characterizing the desaturation rate, $\delta >0$ the upperbound of $|\bm{I}_v|$, 
and $\text{sat}^{\delta}$ a differentiable approximation of the classical saturation function 
defined by $\text{sat}^{\delta}(\bm{x})= \text{min}\left(1,\frac{\delta}{|\bm{x}|}\right)\bm{x}$.
The feedback law of Proposition \ref{prop-velocity-cont} is then applied with  
\begin{IEEEeqnarray}{RCL}
	{\bm \xi}(\bm {\tilde{\rho}}, \bm {\tilde{v}}) &=&  -k_v \bm {\tilde v} -k_i \bm I_v, \nonumber \\
	k_1(\bm k,t) &=& k_{1,0}/(1+\bm k \cdot \bm k_r(t)+\epsilon_1)^2, \nonumber \\
	\lambda(\bm k, t) &=& - \bm \omega_r(t) \cdot \bm k(t),
\end{IEEEeqnarray}
and with $k_v =  5$, $k_i = k^2_v/4$, $k_I=50$, $k_{1,0} = 10$, $\epsilon_1 = 0.01$.  
The feedforward term $\bm{\omega}_r$ is evaluated using the reference acceleration $\dot{\bm v}_r$ rather than the vehicle's acceleration $\dot{\bm v}$ calculated from Newton's equation \eqref{eq:newton0-1} and the model of aerodynamic forces ${\bm F}_a$ used for control design.

The simulated vehicle's equations of motion are given by \eqref{eq:newton0-1}-\eqref{eq:aerodyModelSymmetricBodies}, 
with the aerodynamic coefficients $C_L(\alpha)$ and $C_D(\alpha)$ obtained by interpolating the measurements reported in \cite[p.54]{1971_SAFFEL} 
(see Figure~\ref{fig:ellipticAndMissile}(b)). These coefficients thus differ from the approximating functions \eqref{system:aerodynamicsChaPart} 
used in the control calculation. The values of the parameters involved in these functions are the identified values 
reported previously, i.e. $c_0 =0.1$ and $c_1 = 11.55$. The missile's physical parameters are 
$m = 100 \ [Kg]$ and $(\rho,\Sigma) = (1.292,0.5) \ (\left[ Kg/m^3 \right],[m]^2)$, so that $k_a \approx 0.3 \ \left[ Kg/m \right]$. 
These values are replaced by estimated ones, namely $\hat{k}_a = 0.24$ and $\hat{m} = 80 \ [Kg]$, in the control calculation in order 
to test the control robustness w.r.t. parametric uncertainties. In particular, the vector $\bm{\bar{F}}_p$ in \eqref{Force} is 
calculated 
with $\bm{F}_p({\bm v}_a) = -\hat{k}_a(c_0+2c_1)|\bm{v}_a|\bm{v}_a$. 

The reference velocity $\bm{v}_r(t)$, expressed in Mach numbers ($1 \ Mach = 340 \ \ [m/sec]$), is piece-wise constant on the time interval $[0,40) \ \ [sec]$, and continuously time-varying on the time interval $[40,60) \ \ [sec]$. More precisely:  
\begin{IEEEeqnarray}{RCL}
	\label{vref}
	\bm{v}_r(t) =
	\begin{system}
		0.7 \bm{i}_0  \hspace{1.2cm} 0 \leq &t < 10, \\
		-0.7 \bm{j}_0 \hspace{1cm} 10 \leq &t < 20, \\
		-0.7 \bm{k}_0  \hspace{1cm} 20 \leq &t < 30, \\
		-0.7 \bm{i}_0  \hspace{1cm} 30 \leq &t < 40,
 	\end{system}
\end{IEEEeqnarray}
and $\bm{v}_r(t) =-0.5\sin(t\pi/5) \bm{i}_0 + 0.6\sin(t\pi/10)\bm{j}_0 + 0.6\cos(t\pi/10)\bm{k}_0$ when $40 \leq t < 60$.  
The applied thrust force and angular velocity $\bm{\omega} = (\bm{i},\bm{j},\bm{k})\omega$ are saturated as follows:
\begin{IEEEeqnarray}{RCL}
	0 < &T& < 10 \hat{m}g, \nonumber \\
	|\omega_i| &<& 2\pi, \quad i = \{1,2,3\}. 
\end{IEEEeqnarray}
The initial velocity and attitude are: $\bm{v}(0) = 0.5\bm{i}_0 \ \ [Mach]$,
$\phi_0 = \psi_0 = 0^\circ$, $\theta_0 = -40^\circ$ where $(\phi,\theta,\psi)$ denote standard \emph{roll}, \emph{pitch}, and \emph{yaw} angles as defined in \cite[p. 47]{2004_STENGEL}.

\begin{figure}[!p]
  \vspace{0.4cm}
  \centering
  \small{\input{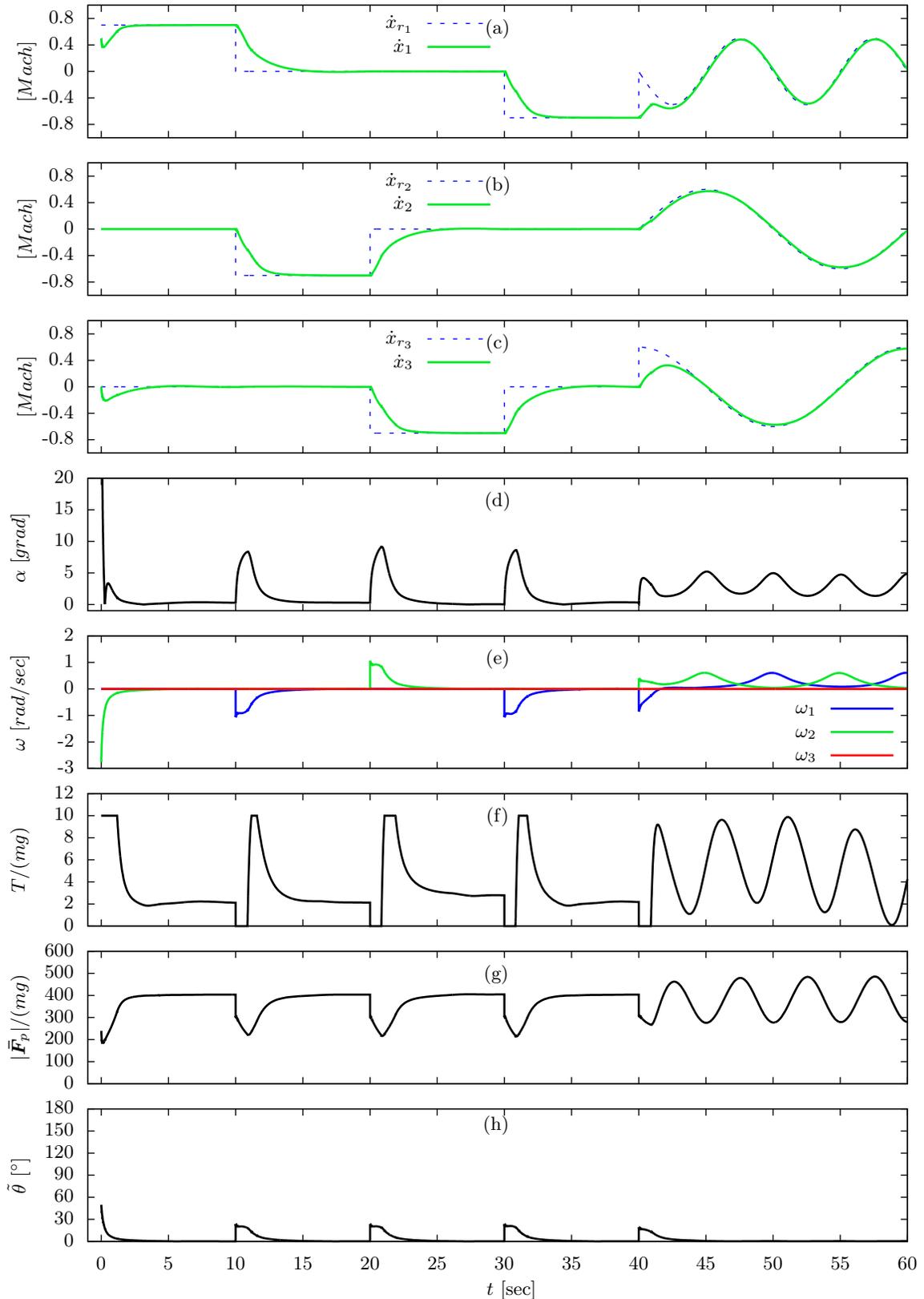}}
  \caption{Simulation of a C-701 trajectory with $\bm{\bar{F}}_p$ in angular control.}
  \label{fig:sim1}
\end{figure}

From top to bottom, Figure~\ref{fig:sim1} shows the time-evolution of the reference velocity $\bm{v}_r = (\bm{i}_0,\bm{j}_0,\bm{k}_0)\dot{x}_r$,
the vehicle's velocity $\bm{v} = (\bm{i}_0,\bm{j}_0,\bm{k}_0)\dot{x}$, the angle of attack $\alpha$, the angular velocity ${\bm \omega}=({\bm \imath},{\bm \jmath},{\bm k})\omega$, 
the applied thrust-to-weight ratio, the norm of the vector $\bm{\bar{F}}_p$ (which has to remain different from zero to ensure the well-posedness of the control solution), and the angle $\tilde{\theta}$ between the thrust direction $\bm{k}$ 
and the reference direction $\bm{k}_r$. There is no wind. 
The initial angle of attack at $t = 0$ is $50^\circ$. The attitude control makes this angle decrease rapidly. 
Sharp discontinuities of the reference velocity at the time instants $t=10,20,30,40\ [sec]$ 
are responsible for the observable transitions and temporarily large angles of attack. Thanks to the integral correction terms resulting from the use of ${\bm I}_v$ in the control law, the velocity error converges to zero when the reference velocity is constant.
On the time interval $[40,60) \ [sec]$, 
despite rapidly varying reference velocities, 
velocity errors are ultimately small, thanks to the combination of pre-compensation and integral correction terms that are present in the control law.

Figure~\ref{fig:sim2} illustrates the improvement brought by the control design proposed in this paper w.r.t. a nonlinear control design that does not take the dependence of the aerodynamic forces upon the vehicle's orientation into account. To this aim, we consider
the velocity control proposed in~\cite{hhms09} for spherical-like vehicles subjected to aerodynamic drag solely. The comparison is facilitated by the fact that this control is basically the same as the one considered in Proposition \ref{prop-velocity-cont} with $\bm{\bar{F}}_a$ used in place of $\bm{\bar{F}}_p$ in the control law. 
Figure~\ref{fig:sim2} shows the evolution of $|\bm{\bar{F}}_a|$ and $\tilde{\theta}$ when applying 
this control,
with the feedforward term $\bm{\omega}_r$ (whose calculation involves $\dot{\bm{F}}_a$) set equal to zero for the sake of simplification.
One can observe from this figure that i) relative variations of the norm of $|\bm{\bar{F}}_a|$ are significantly more pronounced than those of $|\bm{\bar{F}}_p|$ in Figure~\ref{fig:sim1} (a consequence of the dependence of $\bar{\bm{F}}_a$ upon the vehicle's orientation), ii) the amplitude of the orientation error $\tilde{\theta}$ after a discontinuous change of the reference velocity is much more important (an indication of degraded performance), and, even more significantly iii) $\bm{\bar{F}}_a$ crosses zero little after the reference velocity discontinuity occurring at $t=40~[sec]$, with the brisk consequence that the reference direction $\bm{k}_r$, and thus the control law, are not defined at this point (thus leading to an abrupt stop of the simulation).



\begin{figure}[!t]
  \centering
  \small{\input{figure1Sim2.tex}}
  \vspace{0.7cm}
  \caption{Simulation of a C-701 trajectory with $\bm{\bar{F}}_a$ in angular control.}
  \label{fig:sim2}
\end{figure}

\section{Conclusion and perspectives} 
\label{conclusion}
Extension of the thrust direction control paradigm to a class of vehicles with axisymmetric
body shapes has been addressed. Application examples include, e.g., rockets and aerial vehicles using 
annular wings for the production of lift. Specific aerodynamic properties associated with these particular shapes allow 
for the design of nonlinear feedback controllers yielding asymptotic stability in a very large flight envelope.
Further extension of the present approach to vehicles with non-symmetric body shapes (e.g. conventional airplanes) 
is currently investigated in relation to a better understanding of the control limitations induced by the stall phenomenon 
(see e.g. \cite{2012_PUCCI1} for a preliminary study on this latter issue). 
Clearly, the control solution here proposed  calls for a multitude of complementary 
extensions and adaptations before it is implemented on a physical device.  Let us just mention the production of control 
torques allowing for desired angular velocity changes, and the determination of 
corresponding low level control loops that take actuators' physical limitations into account --in relation, for instance, to 
the airspeed dependent control authority associated with the use of flaps and rudders. The addition of actuation degrees of 
freedom via thrust direction ''vectoring'' in order, for instance, to decouple vehicle's attitude control from the constraint of 
thrust direction alignment with the sum of external forces acting on the vehicle, constitutes another extension of the present study.

\section*{Appendix}
We will make use of the following classical vectorial relations:
\begin{equation}
\label{app-vec}
\begin{array}{rl}
\forall \bm x, \bm y, \bm z \in \bm{E}^3, & \quad 
\bm x \cdot (\bm{y} \times \bm z)  = \bm y \cdot (\bm z \times \bm x) \\
\forall \bm x, \bm y, \bm z \in \bm{E}^3, & \quad 
\bm x \times (\bm y \times \bm z) = 
(\bm x \cdot \bm z) \bm y - (\bm x \cdot \bm y) \bm z
\end{array}
\end{equation}

\subsection*{Proof of Proposition \ref{prop-thrust-cont}}
Consider the function $V_0:= 1- \bm k \cdot \bm k_r$ and note that $V_0$ is non-negative and vanishes only when $\bm k= \bm k_r$.
Recall that $\bm \omega_r = \bm k_r \times \dot{\bm k}_r$. By using \eqref{app-vec} and the fact that $\bm k_r$ is a unit vector, one 
deduces that $\dot{\bm k}_r = \bm \omega_r \times \bm k_r$.  The time-derivative of $V_0$ thus satisfies:
\begin{equation}
\label{dV0}
\begin{array}{lcl}
\dot V_0 & = & - \dot{\bm k} \cdot \bm k_r - \bm k \cdot \dot{\bm k}_r \\
 & = & -(\bm \omega \times \bm k) \cdot \bm k_r - \bm k \cdot (\bm \omega_r \times \bm k_r) \\
 & = & (\bm \omega_r - \bm \omega) \cdot (\bm k \times \bm k_r)
\end{array}
\end{equation}
where the last equality follows from \eqref{app-vec}. 
Now, define 
\begin{equation}
\label{app-defV1}
\begin{array}{lcl}
V_1 & := & \displaystyle \frac{\gamma^2(t)}{2}\frac{1- \bm k \cdot \bm k_r}{1 + \bm k \cdot \bm k_r} \\
    & = & \displaystyle \frac{\gamma^2(t)}{2} \frac{1- (\bm k \cdot \bm k_r)^2}{(1 + \bm k \cdot \bm k_r)^2} \\
    & = & \displaystyle \frac{\gamma^2(t)}{2} \frac{|\bm k \times \bm k_r|^2}{(1 + \bm k \cdot \bm k_r)^2}
\end{array}
\end{equation}
where the third equality comes from that $(\bm k \cdot \bm k_r)^2 + |\bm k \times \bm k_r|^2 = 1$, since
$\bm k$ and $\bm k_r$ are unit vectors.
Note also that $V_1= \frac{1}{2} \gamma^2(t) \tan^2(\frac{\tilde{\theta}}{2})$ where $\tilde{\theta}$ is the angle between the vectors $\bm k$ and $\bm k_r$.
One verifies that
\[
\dot V_1 = \gamma(t) \dot \gamma(t) \frac{1- \bm k \cdot \bm k_r}{1 + \bm k \cdot \bm k_r} 
+ \gamma^2(t) \frac{\dot V_0}{(1+ \bm k \cdot \bm k_r)^2}
\]
and it follows from \eqref{dV0} and \eqref{app-defV1} that
\begin{equation}
\label{app-dV1-1}
\begin{array}{lcl}
\dot V_1 & = &  \displaystyle \gamma(t) \dot \gamma(t) \frac{|\bm k \times \bm k_r|^2}{(1 + \bm k \cdot \bm k_r)^2}
+ \gamma^2(t) \frac{(\bm k \times \bm k_r) \cdot (\bm \omega_r - \bm \omega)}{(1+ \bm k \cdot \bm k_r)^2} \\
         & = & \displaystyle \gamma(t) \frac{\bm k \times \bm k_r}{(1+ \bm k \cdot \bm k_r)^2} 
               \bigl( \dot \gamma(t) \, \bm k \times \bm k_r + \gamma(t) (\bm \omega_r - \bm \omega) \bigr)
\end{array}
\end{equation}
Replacing $\bm \omega$ by its expression \eqref{def-omega} yields
\begin{equation}
\label{app-dV1-fin}
\begin{array}{lcl}
\dot V_1 & = & \displaystyle -k_1(\bm k,t) \gamma^2(t) \frac{| \bm k \times \bm k_r|^2}{(1 + \bm k \cdot \bm k_r)^2} \\
         & = & -2 k_1(\bm k,t) V_1
\end{array}
\end{equation}
Since $k_1(\cdot)$ is, by assumption, lower-bounded by a positive scalar, $V_1$ converges exponentially to zero. Exponential
stability of $\bm k = \bm k_r$ then follows from the definition of $V_1$ and the fact that $\gamma(\cdot)$ is lower-bounded 
by a positive scalar.

\subsection*{Proof of Proposition \ref{prop-velocity-cont}}
First, note that in view of Assumption 2 the vector $\bm k_r$ is well defined in a neighborhood of the equilibrium point 
$(\bm{\tilde \rho}, \bm v, \bm k)= (\bm 0, \bm v_r, \bm k_r)$. Then, the term $\gamma(t)$ in 
\eqref{def-omega} is lower-bounded by $\sqrt{c_1}>0$.  Therefore, the feedback law is well defined in a neighborhood of the
equilibrium point.

From \eqref{eq:newton0-1} and \eqref{eq:newFormDynamics}, $\bm F_p - T_p \bm k = \bm F_a - T \bm k$. Therefore 
$\bar{\bm F}_p - T_p \bm k = \bar{\bm F}_a - T \bm k$ and
\[
\begin{array}{lcl}
\bar{\bm F}_p \cdot \bm k & = & T_p - T + \bar{\bm F}_a \cdot k \\
 & = & T_p
\end{array}
\]
From this relation and \eqref{def-kr}, Eq. \eqref{def-mdvt} can be written as follows:
\begin{equation}
\label{app-mdvt}
\begin{array}{lcl}
m \dot{\tilde{\bm v}} & = & |\bar{\bm F}_p| \bm k_r - T_p \bm k + m \bm{\xi}(\tilde{\bm \rho}, \tilde{\bm v}) \\
 & = & |\bar{\bm F}_p| \bm k_r - (\bar{\bm F}_p \cdot \bm k) \bm k + m \bm{\xi}(\tilde{\bm \rho}, \tilde{\bm v}) \\
 & = & |\bar{\bm F}_p| \bm k_r - (|\bar{\bm F}_p| \bm k_r \cdot \bm k) \bm k + m \bm{\xi}(\tilde{\bm \rho}, \tilde{\bm v}) \\
 & = & |\bar{\bm F}_p| ( \bm k_r - (\bm k_r \cdot \bm k) \bm k) + m \bm{\xi}(\tilde{\bm \rho}, \tilde{\bm v}) \\
 & = & |\bar{\bm F}_p| ( \bm k \times (\bm k_r \times \bm k)) + m \bm{\xi}(\tilde{\bm \rho}, \tilde{\bm v})
\end{array}
\end{equation}
where the last equality comes from \eqref{app-vec}. Therefore, along the solutions to the controlled system, the variables 
$\tilde{\bm \rho}$ and $\tilde{\bm v}$ satisfy the following relations:
\begin{equation}
\label{app-drho}
\begin{array}{lcl}
\dot{\bm {\tilde \rho}} & = & \bm f(\bm {\tilde \rho}, \bm {\tilde v}) \\
\dot{\bm {\tilde v}} & = & \bm{\xi}(\bm {\tilde \rho}, \bm {\tilde v}) + \varepsilon
\end{array}
\end{equation}
with the "additive perturbation" $\varepsilon$ defined by
\[
\varepsilon:= \frac{1}{m} |\bar{\bm F}_p| ( \bm k \times (\bm k_r \times \bm k))
\]
From the definition of $\bm \omega$ and Proposition \ref{prop-thrust-cont}, $\bm k$ converges to ${\bm k}_r$ exponentially.
More precisely, from the proof of Proposition \ref{prop-thrust-cont}, the function $V_1$ defined by \eqref{app-defV1} converges to zero 
exponentially. Since $\bm k$ and ${\bm k}_r$ are unit vectors, it follows from \eqref{app-defV1} and the definition of $\gamma$ 
that 
\[
\begin{array}{lcl}
V_1 & \geq & \displaystyle \frac{\gamma^2(t)}{8} |\bm k \times \bm k_r|^2 \\
       & \geq & \displaystyle \frac{|\bar{\bm F}_p|^2}{8} |\bm k \times \bm k_r|^2 \\
       & \geq & \displaystyle \frac{|\bar{\bm F}_p|^2}{8} |\bm k \times (\bm k \times \bm k_r)|^2 \\
       & \geq & \displaystyle \frac{m^2 |\varepsilon|^2}{8}
\end{array}
\]
Therefore, 
\begin{equation}
\label{maj-eps}
\varepsilon \leq \frac{\sqrt{8 V_1}}{m}
\end{equation}
so that $\varepsilon$ also converges to zero exponentially. From Assumption 1 and converse Lyapunov theorems (See, e.g.,  \cite[Section 4.7]{kh02}) there exists a quadratic Lyapunov function 
$V_2(\bm {\tilde \rho}, \bm {\tilde v})$ for System \eqref{def-fullyactuated}, i.e., such that in a neighborhood of 
$(\bm {\tilde \rho}, \bm {\tilde v})=(\bm 0, \bm 0)$, 
\begin{equation}
\label{dv2}
\dot V_2(\bm {\tilde \rho}, \bm {\tilde v}) \leq - k_2 V_2(\bm {\tilde \rho}, \bm {\tilde v})
\end{equation} 
Using the triangular inequality, it follows from \eqref{app-dV1-fin}, \eqref{app-drho}, \eqref{maj-eps}, and \eqref{dv2}  
that the function 
\[
V= \alpha V_1 + V_2
\]
is a Lyapunov function for the controlled system for $\alpha>0$ large enough.

\bibliographystyle{elsarticle-num}        
\bibliography{./bibliography}           

\end{document}